\documentclass[letterpaper,conference,9pt]{IEEEtran}
\IEEEoverridecommandlockouts
\usepackage{pifont}
\newcommand{\cmark}{\ding{51}} 
\newcommand{\xmark}{\ding{55}} 
\usepackage{cite}
\usepackage{amsmath,amssymb,amsfonts}
\usepackage{algorithmic}
\usepackage{graphicx}
\usepackage{textcomp}
\usepackage{xcolor}
\usepackage{algorithmic}
\usepackage{graphicx}
\usepackage{subcaption}
\usepackage{textcomp}
\usepackage{xcolor}
\usepackage{booktabs}
\usepackage{tikz}
\usepackage{graphicx}
\usetikzlibrary{positioning,arrows.meta,calc,matrix,fit}

\usepackage{enumitem}
\usepackage{threeparttable}
\usepackage{multirow}
\usepackage{makecell}
\def\BibTeX{{\rm B\kern-.05em{\sc i\kern-.025em b}\kern-.08em
    T\kern-.1667em\lower.7ex\hbox{E}\kern-.125emX}}

\IEEEoverridecommandlockouts
\IEEEpubid{\makebox[\columnwidth]{979-8-3195-2646-5/26/\$31.00 \copyright2026 IEEE \hfill}
\hspace{\columnsep}\makebox[\columnwidth]{ }}
    
\begin{document}

\title{Toward Deployable Hardware Roots of Trust for Edge AI: Software PUFs and Their Trade-offs\vspace{-0.6em}\\
\thanks{This work was supported in part by the European Horizon MSCA Doctoral Network REACT “Self-AwaRe NEuromorphic ArChiTectures: Security, Reliability and Energy-Efficiency” under Grant 101226463.}
}

\author{\IEEEauthorblockN{Yilin Li and Farhad Merchant}
\IEEEauthorblockA{\textit{Bernoulli Institute and CogniGron}, \textit{University of Groningen}, The Netherlands \\
Email: \{yilin.li, f.a.merchant\}@rug.nl}
\vspace{-25pt}}

\maketitle

\begin{abstract}
Deploying AI at the edge increases the need for security features on existing devices. When physical access to the device is available, storing secret keys in non-volatile memory is vulnerable. Physical unclonable functions (PUFs) address this by exploiting the unique yet repeatable characteristics of each device, but conventional implementations require design-stage modifications, making them inapplicable to a vast number of existing devices already manufactured. Recently, software PUFs have emerged as an alternative approach, leveraging analog characteristics exposed through software-only interfaces to derive unique responses, without post-fabrication changes. Since edge AI hardware is bound to long operational lifespans, this offers clear economic and engineering value. Existing research on software PUFs remains fragmented. It lacks a holistic review of the trade-offs among choices of components used as entropy sources and software methods to expose analog characteristics. 
This gap hinders assessing feasibility, comparing approaches, and identifying open challenges towards a hardware root of trust in existing devices. 
This paper surveys 37 existing software PUF works and compares them against the requirements of on-device model protection: boot-time versus runtime key derivation, and response latency. It analyzes trade-offs, outlines four key challenges that current software PUFs face for edge AI deployment, and highlights promising directions for overcoming barriers to real-world adoption.
\end{abstract}

\vspace{-8pt}

\section{Introduction}

With the proliferation of on-device AI deployment, developers are able to use AI in more scenarios, such as AI in games. However, on-device deployment requires local storage of the model parameters, raising concerns about the security of intellectual property. Even encrypted model weights require the storage of secret keys.
Ren et al.~\cite{ren2024demistify} proposed a method that automatically locates the model, slices the relevant code, and steals the model. Zhou et al.~\cite{zhou2024dynamo} showed that obfuscation can be cracked because the correct model information still needs to be recovered at runtime. Encryption alone does not close the gap: decryption keys must be stored on the device. 
Watermarking~\cite{huang2025themis} can verify ownership after theft but does not prevent copying. What is missing is a protection that binds a model to a specific device, prevents offline copying and redeployment, and works on commodity hardware without modification.


TEE-based protection is an alternative, but it has deployment costs. It depends on hardware extensions or platform support~\cite{moon2025asgard} and on limited secure memory~\cite{abdollahi2025early}, and, as Huang et al.~\cite{huang2025themis} observe, it faces complexity in supporting third-party models and limited adoption in current mobile solutions. Solutions like~\cite{vaswani2023confidential} require hardware extensions in existing devices and thus cannot be deployed to already-fabricated devices. Physical unclonable functions (PUFs) are an attractive solution. The characteristic of not requiring the storage of secret keys makes them suitable candidates for device fingerprinting, key generation, and authentication, enhancing the security of on-device model IP. Despite their security benefits, their widespread adoption is limited because they require invasive, design-stage hardware modifications, making them inapplicable to already-fabricated circuits.

\begin{figure}[!tbp]
    \centering
    \includegraphics[width=0.98\linewidth]{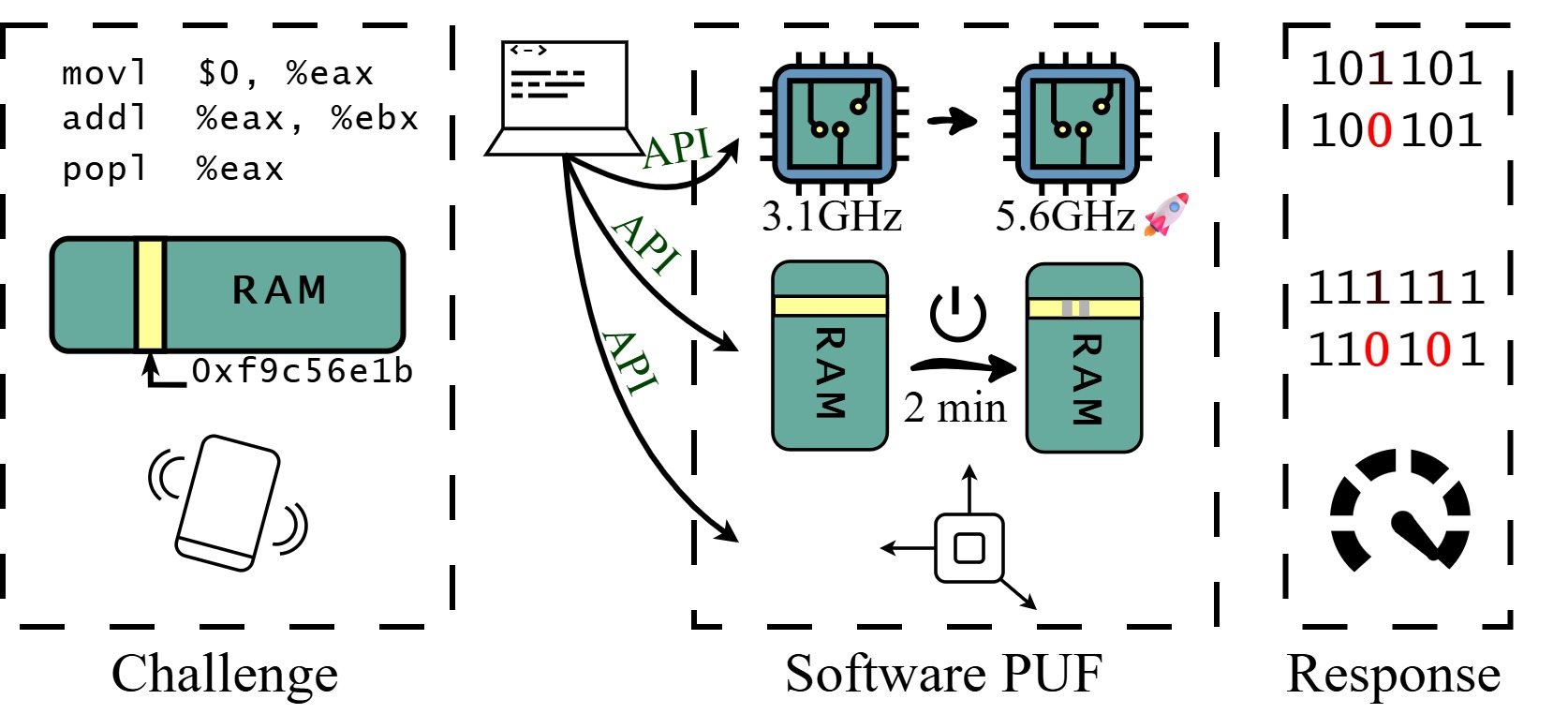}
    \caption{\enspace Challenge and response of software PUFs.}
    \vspace{-14pt}
    \label{fig:CRPs}
\end{figure}

Consequently, enabling PUF functionality on the massive number of existing commercial off-the-shelf (COTS) devices through software-only updates represents a meaningful, zero-hardware-modification security paradigm. 
There have been attempts to exploit static random-access memory (SRAM) \cite{xiao2014bit, liu2017data, dreyer2023espuf, wang2022systematic, schaller2014lightweight, kietzmann2023puf, mispan2022lightweight, bacha2015authenticache}, dynamic random-access memory (DRAM) \cite{schaller2018decay, chen2019intrinsic, xiong2021dram, kim2018dram, miskelly2020fast, najafi2024epuf, chew2025word} and Flash memory~\cite{hasan2024nand, babu2026examination}, which are integral parts of modern computer systems, to implement PUFs. In addition, there are also studies that try to exploit intrinsic process variations in circuitry to implement PUFs \cite{ni2023pi, japa2023processor, japa2024novel, yuan2024dsp, yuan2025overclocking, chen2021spuf, wang2024timing, wang2023novel}. Different methods have been proposed to expose unique device-wise analog characteristics, together with various challenge formats, such as memory addresses \cite{dreyer2023espuf, kietzmann2023puf, mispan2022lightweight, xiong2021dram, najafi2024epuf, chew2025word, hasan2024nand, babu2026examination}, a sequence of instructions \cite{maiti2012novel, ni2023pi, ni2022acf, japa2023processor, japa2024novel}, or even physically shaking the device \cite{fukushima2016sensor, fukushima2019evaluation} and various responses, such as bit flips \cite{chen2019intrinsic, xiong2021dram, hasan2024nand, babu2026examination}, violation errors \cite{najafi2024epuf, chew2025word, japa2024novel, yuan2024dsp, yuan2025overclocking, wang2024timing, wang2023novel}, or sensor readings \cite{fukushima2016sensor, fukushima2019evaluation}, as shown in Fig.~\ref{fig:CRPs}, to implement PUFs. 


Despite this practical value, the scope and limitations of software PUFs remain insufficiently systematized. Widespread adoption is hindered by the difficulty of systematic, fair evaluation across large device populations~\cite{lombardi2025evaluation}. There are three shortcomings in current software PUF studies. 1) The off-the-shelf hardware components that can be exploited have not been systematically summarized. Existing surveys \cite{mcgrath2019puf, chew2025depth} mostly focus on either one class of PUFs using solely one hardware component or the PUFs for specific application purposes, leaving a clear gap for a systematic review on software PUFs. 2) The definition of software PUFs is vague. 
Many works claiming to propose software PUFs \cite{ren2025destruction, tsiokanos2021dta, yang2025rram} fundamentally contradict this classification by requiring invasive, design-stage hardware modifications. 
3) Software PUFs usually exploit the unique and repeatable analog features exposed by errors when hardware is operating in stressed conditions, but the method of inducing abnormal conditions to expose analog characteristics is not well summarized.

With the proliferation of on-device AI deployment, and since hardware is not updated frequently, it is becoming imperative to investigate off-the-shelf solutions to protect model IP, which motivates this work. Despite software PUFs' growing relevance, existing works have proposed isolated implementations, evaluated under different assumptions and metrics, and remain fragmented. A coherent review is lacking to help understand design principles and compare limitations. To address these gaps, this work manually examines prior PUF works that claim to be software PUFs and identifies 37 representative software PUF works. We examine the methods of exposing unique device-wise analog characteristics and summarize them in Table~\ref{tab:software-pufs}.
In summary, the major contributions of this work are as follows: 

\begin{itemize}[leftmargin=*, noitemsep, topsep=0pt]
    \item We provide a precise definition of software PUFs to clearly distinguish them from other types of PUFs.
    \item We present a review of software PUFs, summarize the hardware components exploited, and examine software methods to expose analog characteristics in Section~\ref{software-puf-section}.
    \item We discuss the application of software PUFs in solving model security concerns and the limitations. We identify four key challenges, and derive three future research directions toward secure and practical deployment in Section~\ref{discussion}.
\end{itemize}

The rest of this paper is organized as follows. 
Section~\ref{software-puf-section} defines software PUFs, describes the survey methodology, and reviews existing designs, followed by a cross-category comparison. Section~\ref{discussion} identifies key challenges and discusses promising directions for practical deployment. Finally, Section~\ref{sec:conclusion} concludes the paper.

\section{Background}\label{sec:background}

\subsection{PUFs}

PUFs are hardware primitives that respond to inputs, also called challenges, in a unique, unpredictable, but repeatable way. The characteristic of not requiring the storage of secret keys makes them suitable candidates for device fingerprinting, key generation, and authentication without storing a conventional secret key in nonvolatile memory. There is no single, standalone classification rule that can be applied to all. PUFs can be categorized from different points of view, such as the number of challenge-response pairs (CRPs), the hardware components used, and where the entropy comes from. Depending on the size of CRP space, PUFs can be divided into weak PUFs and strong PUFs~\cite{gao2020physical, dey2021puf}. Weak PUFs provide a limited CRP space and are commonly used for secure boot and cryptographic key generation. Strong PUFs support a large CRP space and are commonly considered for challenge-response authentication protocols. Depending on the hardware used to generate uniqueness, PUFs can be categorized into arbiter PUFs, ring oscillator PUFs, field programmable gate array PUFs, optical PUFs, memory PUFs, etc. 
Based on the entropy source, PUFs are divided into intrinsic and extrinsic, but the term intrinsic is also used for software PUFs in some cases~\cite{gao2020physical, wang2023novel}.


\begin{figure}[t]
\centering
\resizebox{\columnwidth}{!}{%
\begin{tikzpicture}[
    font=\scriptsize,
    node distance=1.2mm,
    box/.style={
        draw,
        rounded corners=2pt,
        align=left,
        text width=41mm,
        inner sep=2pt
    },
    included/.style={box, fill=gray!8},
    excluded/.style={box, fill=gray!22},
    title/.style={
        align=center,
        font=\scriptsize\bfseries,
        text width=41mm,
        inner sep=1pt
    },
    smallnote/.style={
        align=center,
        font=\scriptsize,
        text width=41mm,
        inner sep=1pt
    }
]

\node[title] (lefttitle)  at (0,0)       {Included};
\node[title] (righttitle) at (45.5mm,0)  {Excluded};

\node[included, below=of lefttitle] (sram)
    {{\fontsize{6}{7}\selectfont\bfseries SRAM PUF~\cite{xiao2014bit, liu2017data, dreyer2023espuf, wang2022systematic, schaller2014lightweight, kietzmann2023puf, mispan2022lightweight, bacha2015authenticache}}\\
     Method: bit selection/voltage over-scaling};

\node[included, below=of sram] (dram)
    {{\fontsize{6}{7}\selectfont\bfseries DRAM PUF~\cite{schaller2017intrinsic, tehranipoor2016dram, liu2014trustworthy, rahmati2015probable, xiong2016run, xiong2016practical, sutar2016d, schaller2018decay, chen2019intrinsic, xiong2021dram, kim2018dram, miskelly2020fast, najafi2024epuf, chew2025word}}\\
     Method: rowhammer, refresh reduction, or timing violation};

\node[included, below=of dram] (circuit)
    {{\fontsize{6}{7}\selectfont\bfseries Circuit-intrinsic PUF~\cite{maiti2012novel, ni2023pi, ni2022acf, japa2023processor, japa2024novel, forlin2020g, yuan2024dsp, yuan2025overclocking, chen2021spuf, wang2024timing, wang2023novel}}\\
     Method: timing violation};

\node[included, below=of circuit] (flash)
    {{\fontsize{6}{7}\selectfont\bfseries Program-disturb Flash PUF~\cite{hasan2024nand, babu2026examination}}\\
     Method: repeated page programming};

\node[included, below=of flash] (sensor)
    {{\fontsize{6}{7}\selectfont\bfseries Sensor PUF~\cite{fukushima2016sensor, fukushima2019evaluation}}\\
     Method: accelerometer/gyroscope access};

\node[excluded, below=of righttitle] (isa)
    {{\fontsize{6}{7}\selectfont\bfseries Processor-/ISA-modified PUF~\cite{tsiokanos2021dta, aditya2026methodology}}\\
     Requires ISA extension or hardware logic modification.};

\node[excluded, below=of isa] (dedicated)
    {{\fontsize{6}{7}\selectfont\bfseries Dedicated-circuitry PUF~\cite{tashdid2026interpuf, younes2025dcdc}}\\
     Requires dedicated RO, arbiter, or delay circuitry.};

\node[excluded, below=of dedicated] (memorymod)
    {{\fontsize{6}{7}\selectfont\bfseries Modified memory PUF~\cite{ni2024s2ram}}\\
     Requires memory-array or peripheral-circuit modification.};

\node[excluded, below=of memorymod] (nvm)
    {{\fontsize{6}{7}\selectfont\bfseries RRAM, FeRAM, and MRAM PUF~\cite{rajendran2024harnessing}}\\
     Non-volatile memory is not widely available in COTS devices yet.};

\node[smallnote, below=0.8mm of sensor] (leftnote)
    {Zero hardware modification:\\deployable on COTS devices};

\node[smallnote, below=0.8mm of nvm] (rightnote)
    {Hardware modification required:\\not deployable by software update alone};

\coordinate (topbound) at ($(lefttitle.east)!0.5!(righttitle.west)$);
\path let \p1=(topbound), \p2=(leftnote.south), \p3=(rightnote.south),
          \n1={min(\y2,\y3)}
      in coordinate (botbound) at (\x1,\n1);

\node[
    align=center,
    font=\scriptsize\bfseries,
    anchor=south,
    inner sep=1pt
] (boundarylabel) at (topbound)
    {zero-HW-modification\\boundary};

\draw[dashed, thick] (boundarylabel.south) -- (botbound);

\end{tikzpicture}%
}
\caption{Scope boundary of software PUFs in commodity devices. The left side follows categories summarized in Table~\ref{tab:software-pufs}; the right side shows the designs that fall outside the software-PUF scope because they require hardware modification.}
\vspace{-10pt}
\label{fig:software-puf-boundary}
\end{figure}
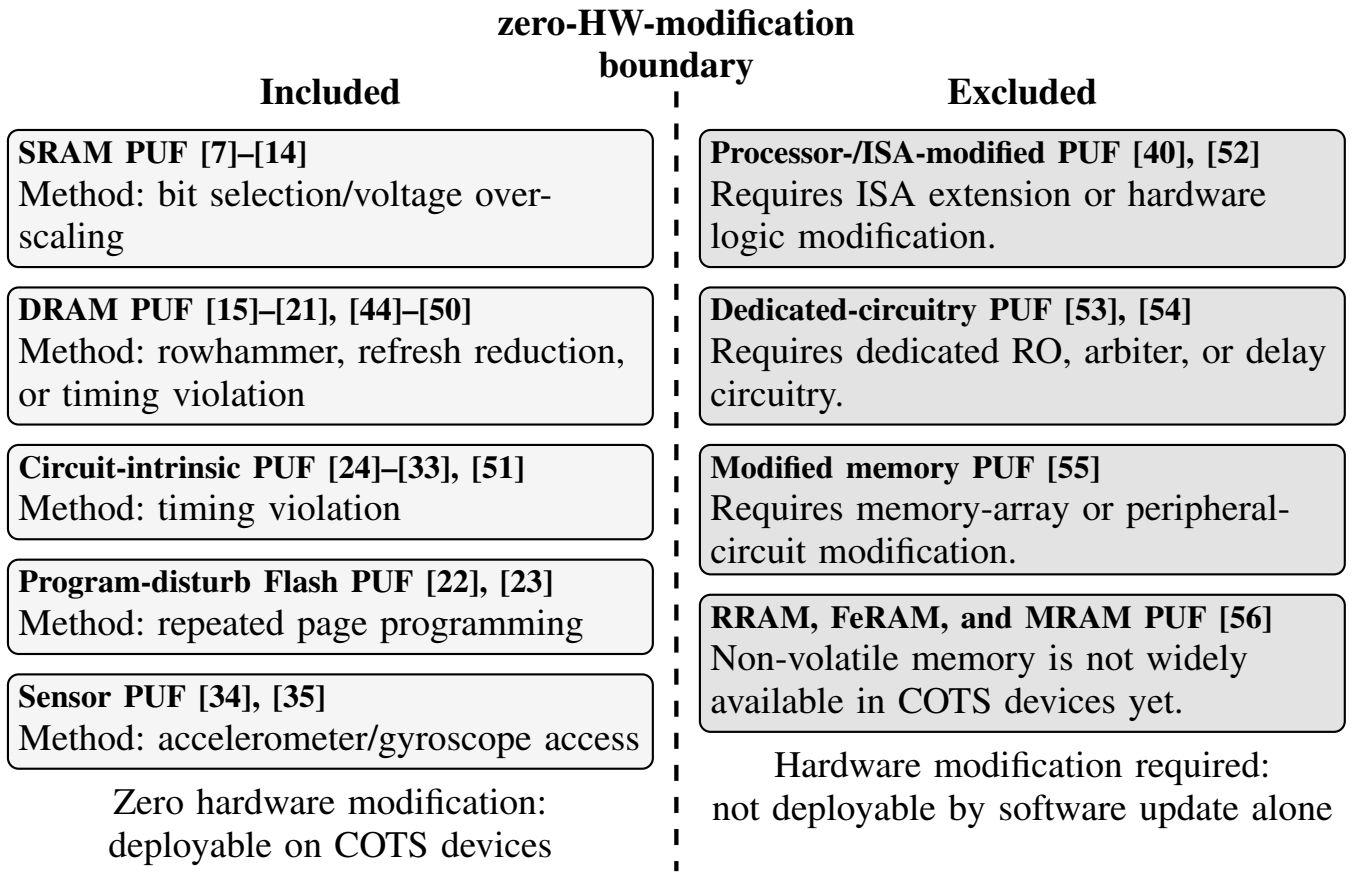

\subsection{On-Device AI Deployment}

On-device AI stores model weights locally, instead of sending input to the cloud and waiting for response from cloud. Bringing AI to the edge eases the burden of cloud servers and preserves user privacy. One example of on-device AI is the on-device voice assistant on smartphones, which processes speech locally without uploading audio. However, deploying AI at the edge also means that the attacker has easier access to the model weights, raising intellectual property security concerns.

\begin{table*}[tbp]
\vspace{-0.5em}
\centering
\resizebox{\linewidth}{!}{
\begin{threeparttable}
\caption{Summary of software PUFs.}
\label{tab:software-pufs}
\setlength{\tabcolsep}{2pt}
\begin{tabular}{c|lcccccccc}
\toprule
\multicolumn{2}{c}{Category} & Research & Challenge & Response & Method & Uniqueness (\%) & Reliability (\%) & \makecell[c]{Runtime\\accessibility} & \makecell[c]{Fuzzy\\extractor} \\ 
\midrule
\multirow{8}{*}{\rotatebox{90}{SRAM PUF}} & \multirow{7}{*}{start-up} & Xiao et al.~\cite{xiao2014bit} & \multirow{7}{*}{\makecell[c]{memory\\address}} & \multirow{7}{*}{start-up values} & neighborhood bit selection & $\overline{\mathrm{HD}}_{\mathrm{inter}} \simeq 49.95*$ & $\mathrm{HD}_{\mathrm{intra}} \leq 3.13$* & \xmark & \xmark \\ 
 & & Liu et al.~\cite{liu2017data} &  &  & remanence-based selection & $\overline{\mathrm{HD}}_{\mathrm{inter}} = 49.35$ & $\mathrm{HD}_{\mathrm{intra}} = 0$ & \xmark & \xmark \\
 & & Wang et al.~\cite{wang2022systematic} &  &  & ramp-rate + data-retention & --& -- & \xmark & \xmark \\
 & & ESPuF~\cite{dreyer2023espuf} &  &  & statistics-based bit selection & $\overline{\mathrm{HD}}_{\mathrm{inter}} = 51.2$ & -- & \xmark & \xmark \\
 & & Schaller et al.~\cite{schaller2014lightweight} &  &  & -- & $\overline{\mathrm{HD}}_{\mathrm{inter}} \simeq 47.53*$ & $\mathrm{HD}_{\mathrm{intra}} \leq 9.61$ & \xmark & \cmark \\
 & & Kietzmann et al.~\cite{kietzmann2023puf} &  &  & -- & $\overline{\mathrm{HD}}_{\mathrm{inter}} \simeq 48$ & $\mathrm{HD}_{\mathrm{intra}} \lesssim 7*$ & \xmark & \cmark \\
 & & Mispan et al. ~\cite{mispan2022lightweight} &  &  & randomly selected bits & $\overline{\mathrm{HD}}_{\mathrm{inter}} = 42.64$ & $\mathrm{HD}_{\mathrm{intra}} \leq 6.71$ & \xmark & \xmark \\
 \cmidrule{2-10}
 & violation & Authenticache~\cite{bacha2015authenticache} & memory address & cache errors & voltage over-scaling & $\overline{\mathrm{HD}}_{\mathrm{inter}} = 49$ & $\mathrm{HD}_{\mathrm{intra}} \leq 6$ & \cmark & \cmark \\
\midrule
\multirow{14}{*}{\rotatebox{90}{DRAM PUF}} 
 & rowhammer & Schaller et al.~\cite{schaller2017intrinsic} & memory address & flipped values & rowhammer & $J_{\mathrm{inter}} \lesssim 4$ & $J_{\mathrm{intra}} \geq 94.54$ & \cmark & \xmark\\
 \cmidrule{2-10}
 & start-up & Tehranipoor et al.~\cite{tehranipoor2016dram} & memory address & start-up values & bit selection & $\overline{\mathrm{HD}}_{\mathrm{inter}} = 49.37$ & $\overline{\mathrm{HD}}_{\mathrm{intra}} \simeq 2*$ & \xmark & \xmark \\
 \cmidrule{2-10}
 & \multirow{8}{*}{retention} & Liu et al.~\cite{liu2014trustworthy} & \multirow{8}{*}{\makecell[c]{memory\\address}} & \multirow{8}{*}{decayed values} & \multirow{8}{*}{decreased refresh rate} & $\mathrm{HD}_{\mathrm{inter}} \simeq 34$ & $\mathrm{HD}_{\mathrm{intra}} \lesssim 5.2$ & \xmark & \cmark \\
 & & Rahmati et al.~\cite{rahmati2015probable} &  &  &  & -- & $\overline{\mathrm{HD}}_{\mathrm{intra}} \simeq 1$ & \xmark & \xmark \\
 & & Xiong et al.~\cite{xiong2016run} & &   &  & $J_{\mathrm{inter}} \leq 4.05$ & $J_{\mathrm{intra}} \geq 34.84$ & \cmark & \cmark \\
 & & Xiong et al.~\cite{xiong2016practical} &  &  &  & $\overline{\mathrm{HD}}_{\mathrm{inter}} \simeq 5*$ & $\mathrm{HD}_{\mathrm{intra}} \leq 3.86$ & \cmark & \cmark \\
 & & D-PUF~\cite{sutar2016d} &  &  &  & $\overline{\mathrm{HD}}_{\mathrm{inter}} = 0.034*$ & $\overline{\mathrm{HD}}_{\mathrm{intra}} \simeq 0.005*$ & -- & \cmark \\
 & & Schaller et al.~\cite{schaller2018decay} &  &  &  & $J_{\mathrm{inter}} \leq 4.1$ & $J_{\mathrm{intra}} \geq 38.5$ & \cmark & \cmark \\
 & & Chen et al.~\cite{chen2019intrinsic} &  &  &  & $J_{\mathrm{inter}} \simeq 0.015$ & -- & \cmark & \cmark \\
 & & Xiong et al.~\cite{xiong2021dram} &  &  &  & $J_{\mathrm{inter}} \leq 4.1$ & $J_{\mathrm{intra}} \geq 34.84$ & \cmark & \cmark \\
 \cmidrule{2-10}
 & \multirow{4}{*}{latency} & Kim et al.~\cite{kim2018dram} & \multirow{4}{*}{\makecell[c]{memory\\address}} & \multirow{4}{*}{latency failures} & \multirow{4}{*}{timing violation} & $J_{\mathrm{inter}} \leq 25$ & $J_{\mathrm{intra}} \geq 65$  & \cmark & \xmark \\
 & & Miskelly et al.~\cite{miskelly2020fast} &  &  &  & $\overline{\mathrm{HD}}_{\mathrm{inter}} = 47.95$ & $\overline{\mathrm{HD}}_{\mathrm{intra}} = 6.41*$ & \cmark & \cmark \\
 & & EPUF~\cite{najafi2024epuf} &  &  &  & $\overline{\mathrm{HD}}_{\mathrm{inter}} = 47.79$ & $\overline{\mathrm{HD}}_{\mathrm{intra}} \simeq 0.03*$ & \cmark & \cmark \\
 & & Chew et al.~\cite{chew2025word} &  &  &  & $J_{\mathrm{inter}} = 49.31*$ & $\mathrm{HD}_{\mathrm{intra}} \leq 3.125*$ & -- & \cmark \\
\midrule
\multicolumn{2}{l}{\multirow{2}{*}{program-disturb Flash PUF}} & Hasan et al.~\cite{hasan2024nand} & \multirow{2}{*}{\makecell[c]{page address}} & \multirow{2}{*}{flipped bits} & \multirow{2}{*}{\makecell[c]{repeated page programming}} & -- & -- & \cmark & -- \\
 \multicolumn{2}{l}{\multirow{2}{*}{}} & Babu et al.~\cite{babu2026examination} &  &  &  & $\overline{\mathrm{HD}}_{\mathrm{inter}}=42.01*$ & $\overline{\mathrm{HD}}_{\mathrm{intra}}=12.63*$ & \cmark & -- \\
\midrule
\multirow{13}{*}{\rotatebox{90}{Circuit-intrinsic PUF}} & \multirow{5}{*}{processor} & Maiti et al.~\cite{maiti2012novel} & \multirow{5}{*}{\makecell[c]{instructions}} & \multirow{5}{*}{timing errors} & overclocking & $\overline{\mathrm{HD}}_{\mathrm{inter}}=37.54*$ & $\overline{\mathrm{HD}}_{\mathrm{intra}} = 2.086$ & -- & \xmark \\
 & & PI PUF~\cite{ni2023pi} &  &  & overclocking & $\overline{\mathrm{HD}}_{\mathrm{inter}}=50.93$ & $\overline{\mathrm{HD}}_{\mathrm{intra}} = 1.13$ & -- & \cmark \\ 
 & & Ni et al.~\cite{ni2022acf} &  &  & voltage over-scaling & $\overline{\mathrm{HD}}_{\mathrm{inter}}=50.10$ & $\overline{\mathrm{HD}}_{\mathrm{intra}} \simeq 9.818*$ & -- & \cmark \\
 & & Japa et al.~\cite{japa2023processor} &  &  & voltage over-scaling & $\overline{\mathrm{HD}}_{\mathrm{inter}}=30.66 $ & $\overline{\mathrm{HD}}_{\mathrm{intra}} = 7.988$ & -- & -- \\
 & & Japa et al.~\cite{japa2024novel} &  &  & voltage over-scaling & $\overline{\mathrm{HD}}_{\mathrm{inter}} = 46.45*$ & $\overline{\mathrm{HD}}_{\mathrm{intra}} = 10.5*$ & -- & -- \\
 \cmidrule{2-10}
 & GPU & G-PUF~\cite{forlin2020g} & matrix multiply & timing errors & overclocking & $\overline{\mathrm{HD}}_{\mathrm{inter}} = 55.61$ & $\overline{\mathrm{HD}}_{\mathrm{intra}}=9.91$ & \cmark & -- \\
 \cmidrule{2-10}
 & \multirow{5}{*}{others} & DSP-PUF~\cite{yuan2024dsp} & \multirow{5}{*}{data stream} & \multirow{5}{*}{timing errors} & \multirow{5}{*}{overclocking} & -- & $\mathrm{HD}_{\mathrm{intra}} \leq 11.99$ & -- & -- \\  
 & & Yuan et al.~\cite{yuan2025overclocking} &  &  &  & $\overline{\mathrm{HD}}_{\mathrm{inter}} = 48.87$ & $\mathrm{HD}_{\mathrm{intra}} \leq 33.34$ & -- & -- \\ 
 & & SPUF~\cite{chen2021spuf} &  &  &  & $\overline{\mathrm{HD}}_{\mathrm{inter}} = 50.01$ & -- & -- & -- \\
 & & TVS-PUF~\cite{wang2024timing} &  &  &  & $\overline{\mathrm{HD}}_{\mathrm{inter}} = 49.93$ & $\overline{\mathrm{HD}}_{\mathrm{intra}} = 2.61$ & -- & --\\
 & & FAS-PUF~\cite{wang2023novel} &  &  &  & $\overline{\mathrm{HD}}_{\mathrm{inter}} = 49.54$ & $\overline{\mathrm{HD}}_{\mathrm{intra}} = 4.67$ & -- & -- \\
\midrule
\multicolumn{2}{l}{\multirow{2}{*}{Sensor PUF}} & Fukushima et al.~\cite{fukushima2016sensor} & \multirow{2}{*}{\makecell[c]{shaking devices}} & \multirow{2}{*}{values from sensor} & accelerometer/gyroscope & $\overline{\mathrm{HD}}_{\mathrm{inter}} = 49.498*$ & -- & \cmark & \cmark \\
 \multicolumn{2}{l}{\multirow{2}{*}{}} & Fukushima et al.~\cite{fukushima2019evaluation} &  &  & gyroscope & $\overline{\mathrm{HD}}_{\mathrm{inter}} \simeq 49.83*$ & -- & \cmark & \cmark \\
\bottomrule
\end{tabular}
\vspace{0pt}
\begin{tablenotes}
      \footnotesize
      \item $\mathrm{HD}$ and $J$ denote the fractional Hamming distance and Jaccard similarity, respectively. $\overline{\mathrm{HD}}$ and $\overline{J}$ denote their mean values. Intra-device measurements refer to evaluations within the same device (reliability), while inter-device measurements refer to evaluations across different devices (uniqueness). 
      \item For Hamming distance, higher inter-device and lower intra-device values are preferable; for Jaccard similarity, lower inter-device and higher intra-device values are preferable.
      \item * Values are (re)calculated from the reported data rather than explicitly provided by the original authors.
      \item A dash indicates that the corresponding information was not reported or could not be determined from the original publication.
    \end{tablenotes}
\end{threeparttable}
}
\vspace{-10pt}
\end{table*}

\section{Software PUFs}\label{software-puf-section}

\subsection{Definition}

There are several works \cite{ren2025destruction, tsiokanos2021dta, yang2025rram} that claim to implement software PUFs while still requiring hardware logic modifications, which causes confusion. To clarify, we define software PUFs as \textbf{PUFs that exploit existing hardware components not originally designed for PUF functionality, thereby enabling PUF capabilities where none previously existed, without requiring design-stage modification to the hardware logic.} The zero-hardware-modification requirement is the key boundary of this survey. A design that can be realized purely through software on already-fabricated hardware is deployable on existing commodity devices. In contrast, a design that requires even minor hardware additions, modified logic, ISA extensions, dedicated PUF circuits, or special power-management circuitry loses this retrofitting property. Such designs may still be useful PUF architectures, but they should not be classified as software PUFs under the COTS-device deployment model.

The proposed taxonomy does not replace conventional PUF classifications. Instead, it introduces an additional classification dimension based on accessibility. A software PUF is defined by how the response is obtained, through software-level interfaces without design-stage hardware changes, whereas SRAM, DRAM, and sensor PUFs describe where the entropy originates. Accordingly, we organize prior works along two axes: entropy source and software-controlled exposure method, as in Table~\ref{tab:software-pufs}. The software may involve privileged system software, firmware, device drivers, or kernel modules, provided that the response-exposure mechanism requires no design-stage modification to the underlying hardware.

\subsection{Survey Methodology and Scope}

We include a work as a software PUF only if it satisfies the following criteria: 

\begin{itemize}[leftmargin=*, noitemsep, topsep=0pt]
    \item the entropy source comes from existing hardware components that were not originally designed for PUF functionality; 
    \item the PUF response can be exposed through software-controlled operations, such as memory access, timing-parameter adjustment, voltage or frequency stress, or sensor reading; and 
    \item the design does not require design-stage modification of hardware logic, additional dedicated PUF circuitry, or changes to the processor microarchitecture.
\end{itemize}

We exclude works that require custom hardware changes. We also distinguish software PUFs from intrinsic PUFs: intrinsic PUFs refer to the source of entropy, whereas software PUFs refer to the implementation and accessibility model. Fig.~\ref{fig:software-puf-boundary} summarizes the key boundary used in this survey. To construct the survey, we used a recursive backward and forward snowballing-based manual review, seeding from an early processor-intrinsic software PUF~\cite{maiti2012novel}, an SRAM PUF~\cite{wang2022systematic} and a recent DRAM latency PUF~\cite{chew2025word}. We examined prior work claiming or resembling software PUFs and categorized them according to their entropy source. We also extracted uniqueness and reliability metrics when reported, and converted the reported numbers, where possible, to fractional inter- and intra-Hamming distances (HD) or Jaccard similarity ($J$) to facilitate comparison. Recalculated values are marked with an asterisk in Table~\ref{tab:software-pufs}. 

\subsection{Overview}

\begin{figure}[!tbp]
    \centering
    \includegraphics[width=1\linewidth]{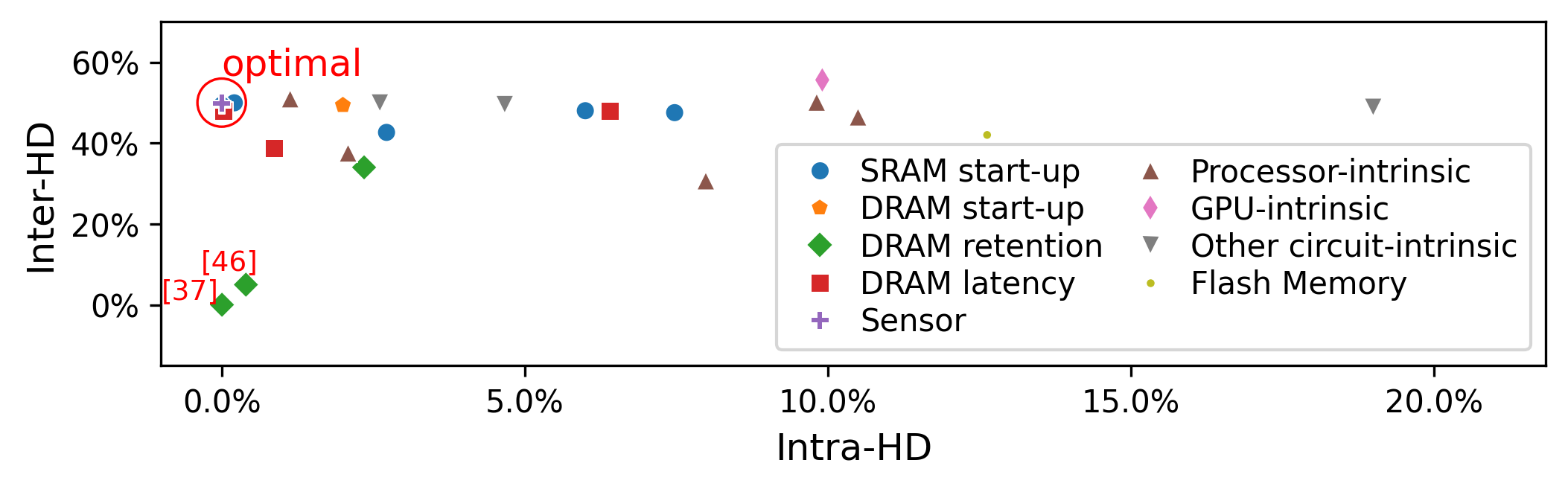}
    \vspace{-16pt}
    \caption{\enspace Overview of the uniqueness and reliability of software PUFs, evaluated using mean fractional inter- and intra-HDs. Not all entries in Table~\ref{tab:software-pufs} report HDs. Some values were derived from the reported data, while entries without sufficient information were omitted.}
    \label{fig:software-PUF-HDs}
    \vspace{-14pt}
\end{figure}

Based on the above inclusion criteria, we identify 37 representative software PUF works, as shown in Table~\ref{tab:software-pufs}, categorized according to the component from which the entropy source comes and the methods used to expose analog characteristics. We also manually checked the performance reported in these works and plotted Fig.~\ref{fig:software-PUF-HDs}. Since evaluation settings differ across studies, the reported values should be interpreted as comparative indicators rather than as results from a controlled benchmark. We derive two observations based on the analysis. 

\textbf{1) The measurement metrics are inconsistent, though among them uniqueness and reliability, measured by the inter- and intra-HD and $J$, are commonly used}. 
We also observed that the HD may not be a reliable metric for evaluating uniqueness and reliability under certain conditions, as illustrated in Fig.~\ref{fig:software-PUF-HDs}. 
For sparse software PUF responses, HD can be misleading because most positions are shared zeros, as in~\cite{xiong2016practical} and~\cite{sutar2016d}. $J$ is more informative in such cases because it measures the overlap of active response positions.


\textbf{2) Most software PUFs require an enrollment stage to determine hyperparameters}. For example, some SRAM PUFs use the enrollment stage to select stable bits, while some DRAM PUFs and processor-intrinsic PUFs use this stage to select appropriate operating frequencies.

\subsection{SRAM PUFs}

Cortez et al. \cite{cortez2012modeling} observed that SRAM cells are inherently imbalanced, meaning that different cells exhibit varying biases toward preferred start-up values, as in Fig.~\ref{fig:sram-dram-principles}%
(\subref{fig:startup-sram-puf}). 
The SRAM cell start-up values exhibit repeatable and device-specific behavior and can serve as PUF responses when stable cells are carefully selected.
Xiao et al. \cite{xiao2014bit} observed that power-up stability is related to neighboring cells and proposed a neighborhood-based bit selection method. Although the authors claimed that knowledge of the physical hardware SRAM structure is needed, this work requires no hardware modification. 
Liu et al. \cite{liu2017data} adopted a data remanence-based approach to rapidly select stable bits, suitable for high-volume production flow. The idea is to first write all SRAM cells with 1, then power off the SRAM for a small period of time, and finally select the cells that flipped bits as stable cells. 
The only factor to be determined is the power down time. Wang et al. \cite{wang2018exploiting} investigated the root cause of SRAM cell skewness. Their experimental evidence suggests that power-up values of SRAM cells are strongly dependent on the ramp rate and direction of power supply. 
Based on this observation, they proposed to use three different tests under different power supply and combine the results to identify stable cells. 
Wang et al.~\cite{wang2022systematic} later proposed a two-stage bit-selection method 
to identify strongly biased cells.
However, one essential constraint of SRAM PUFs based on start-up values is that they are not accessible during runtime. There are also arguments \cite{kietzmann2023puf} that such bit selection adds complexity to enrollment and conflicts with the limited memory resource on Internet of Things (IoT) devices. Unlike the other designs, Authenticache \cite{bacha2015authenticache} is a runtime-accessible SRAM PUF using on-chip error correction logic.

\subsection{DRAM PUFs}

\begin{figure}[t]
\centering

\begin{subfigure}[t]{0.457\columnwidth}
\centering
\resizebox{\linewidth}{!}{%
\begin{tikzpicture}[
    font=\small,
    >=stealth,
    arrow/.style={->, thick},
    cell/.style={
        draw,
        minimum width=5.5mm,
        minimum height=5.5mm,
        inner sep=0pt
    },
    offcell/.style={
        cell,
        fill=gray!45
    },
    one/.style={
        cell,
        fill=black,
        text=white
    },
    zero/.style={
        cell,
        fill=white
    }
]


\node[offcell] at (-5.4,1.7) {};
\node[offcell] at (-4.8,1.7) {};

\node[offcell] at (-5.4,1.1) {};
\node[offcell] at (-4.8,1.1) {};

\node[offcell] at (-4.2,1.7) {};
\node[offcell] at (-3.6,1.7) {};

\node[offcell] at (-4.2,1.1) {};
\node[offcell] at (-3.6,1.1) {};

\node[one]  at (-1.3,1.7) {1};
\node[zero] at (-0.7,1.7) {0};

\node[zero] at (-1.3,1.1) {0};
\node[one]  at (-0.7,1.1) {1};

\node[zero] at (-0.1,1.7) {0};
\node[zero] at (0.5,1.7) {0};

\node[one]  at (-0.1,1.1) {1};
\node[zero] at (0.5,1.1) {0};

\draw[arrow]
    (-3.25,1.4) --
    node[above] {power-up}
    (-1.65,1.4);

\node[align=center] at (-4.5,0.5)
    {powered off};

\node[align=center] at (-0.5,0.5)
    {start-up values};

\end{tikzpicture}%
}
\caption{Start-up SRAM PUF}
\label{fig:startup-sram-puf}
\end{subfigure}%
\hfill
\begin{subfigure}[t]{0.475\columnwidth}
\centering
\resizebox{\linewidth}{!}{%
\begin{tikzpicture}[
    font=\small,
    >=stealth,
    arrow/.style={->, thick},
    cell/.style={
        draw,
        minimum width=5.5mm,
        minimum height=5.5mm,
        inner sep=0pt
    },
    offcell/.style={
        cell,
        fill=gray!45
    },
    one/.style={
        cell,
        fill=black,
        text=white
    },
    zero/.style={
        cell,
        fill=white
    }
]


\node[one] at (-0.5,1.7) {1};
\node[one] at (0.1,1.7) {1};

\node[one] at (-0.5,1.1) {1};
\node[one] at (0.1,1.1) {1};

\node[one] at (0.7,1.7) {1};
\node[one] at (1.3,1.7) {1};

\node[one] at (0.7,1.1) {1};
\node[one] at (1.3,1.1) {1};

\node[one]  at (3.6,1.7) {1};
\node[zero] at (4.2,1.7) {0};

\node[zero] at (3.6,1.1) {0};
\node[one]  at (4.2,1.1) {1};

\node[one]  at (4.8,1.7) {1};
\node[zero] at (5.4,1.7) {0};

\node[zero] at (4.8,1.1) {0};
\node[one]  at (5.4,1.1) {1};

\draw[arrow]
    (1.65,1.4) --
    node[above, align=center] {refresh off\\and wait}
    (3.25,1.4);

\node[align=center] at (0.4,0.5)
    {written pattern};

\node[align=center] at (4.4,0.5)
    {decayed values};

\end{tikzpicture}%
}
\caption{Retention DRAM PUF}
\label{fig:retention-dram-puf}
\end{subfigure}
\caption{Principles of start-up SRAM and retention DRAM PUFs.
Gray cells denote the undefined SRAM state while powered off;
black and white cells denote logic 1 and logic 0.}
\vspace{-14pt}
\label{fig:sram-dram-principles}
\end{figure}
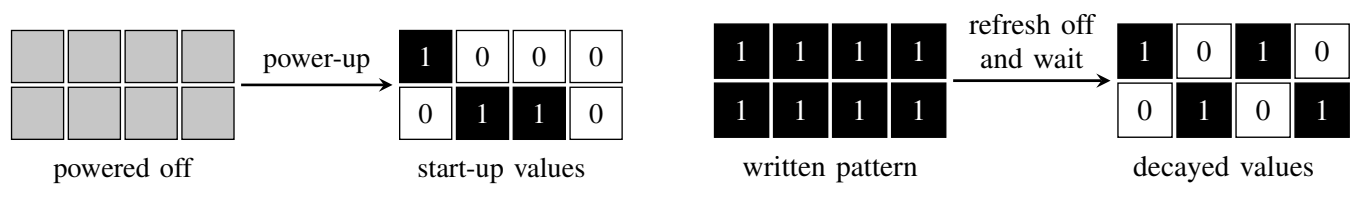

There are two major DRAM PUF categories: retention PUFs \cite{liu2014trustworthy, rahmati2015probable, xiong2016run, xiong2016practical, sutar2016d, schaller2018decay, chen2019intrinsic, xiong2021dram} and latency PUFs \cite{kim2018dram, miskelly2020fast, najafi2024epuf, chew2025word}, as shown in Table~\ref{tab:software-pufs}. Retention PUFs exploit the fact that a DRAM cell loses its stored value if it is not refreshed in time, as in Fig.~\ref{fig:sram-dram-principles}%
(\subref{fig:retention-dram-puf}). The maximum duration for which data can be reliably retained in a DRAM cell varies from cell to cell due to process variations. Liu et al. \cite{liu2014trustworthy} proposed a DRAM PUF: by setting the refreshing period to 120 seconds, they obtained a 4080-bit PUF response. Rahmati et al. \cite{rahmati2015probable} observed that approximate computing utilizes an idea similar to retention PUFs. However, \cite{liu2014trustworthy} and \cite{rahmati2015probable} exist only under experimental conditions and are not accessible at runtime. 
Xiong et al. \cite{xiong2016run} observed that even on platforms where the DRAM refresh rate can only be controlled at the granularity of entire DRAM, the memory can be selectively refreshed implicitly by reading or writing rows. Combining selective refreshing with the memory ballooning technique developed for virtual machines, they implemented a kernel module to support runtime queries. 

Retention DRAM PUFs have a major practical limitation of minute-scale latency. Kim et al. \cite{kim2018dram} proposed latency PUFs, exploiting the DRAM latency failure when operating at reduced timing parameters, reducing the DRAM PUF response time from minutes to milliseconds. Miskelly et al. \cite{miskelly2020fast} further reduced the latency of DRAM PUF by redesigning the method and postponing the processing of the raw PUF response to when the memory has been released. In addition to that, Schaller et al. \cite{schaller2017intrinsic} observed that the rowhammer effect, a vulnerability in DRAM where repeated access to a memory row may cause nearby row bit flips, induces unique device-specific characteristics that can be used for PUF. Tehranipoor et al. \cite{tehranipoor2016dram} observed that some DRAM devices exhibit similar behavior, where memory cells have preferred start-up values that can be exploited as a PUF.

\subsection{Circuit-intrinsic PUFs}
\vspace{-4pt}

\subsubsection{Processor-intrinsic PUFs}

Maiti et al. \cite{maiti2012novel} observed that when the microcontroller unit (MCU) operates under overclocking conditions, device-specific variations introduced during the manufacturing process manifest themselves as differences in circuit delays. By carefully selecting the timing parameters, the MCU remains functionally correct for most operations, while errors occur only during timing-critical computations. This device-specific variation can be characterized by analyzing computation errors under these carefully selected timing parameters.
PI-PUF~\cite{ni2023pi} proposed a multi-round instruction obfuscation mechanism and used the Gray code to improve the uniqueness and randomness of the PUF. Unlike other processor-intrinsic PUFs, Ni et al.~\cite{ni2022acf} exploited low-voltage-induced timing violations to generate responses. However, PI-PUF~\cite{ni2023pi} and Ni et al.~\cite{ni2022acf} did not mention how clock frequency or voltage is safely manipulated in real circuits. To date, no processor-intrinsic PUF has been demonstrated on COTS processors.

There are some processor-based PUF designs \cite{tsiokanos2021dta, zhao2024high, kong2014pufatt} that do not qualify as software PUFs because they require hardware additions. DTA-PUF \cite{tsiokanos2021dta} outperforms the design of Maiti et al. in response bit generation speed by systematically exploring the dynamic timing behavior of logic. 
However, the instruction set architecture was extended to support clock frequency tuning. Zhao et al. \cite{zhao2024high} and Kong et al. \cite{kong2014pufatt} are excluded because they require hardware additions of arbiters. 

\subsubsection{Other circuit-intrinsic PUFs}

\begin{table}[bp]
\vspace{-2pt}
\centering
\resizebox{\linewidth}{!}{
\begin{threeparttable}
\caption{Comparison of different types of software PUFs.}

\begin{tabular}{lcl}
\toprule
PUF & \makecell[c]{Latency} & \makecell[c]{Limitation} \\ 
\midrule
start-up SRAM & ms–s & limited runtime accessibility \\
violation SRAM & $\simeq$ 125ms & voltage-parameter tuning \\
\midrule
rowhammer DRAM & $\simeq$ 1-2mins & aging and limited entropy \\
start-up DRAM & minutes & limited runtime accessibility\\
retention DRAM & $\simeq$ 1-6mins & minute-scale response time\\
latency DRAM & $\simeq$ 3-90ms & \makecell[l]{fine-grained timing control} \\
\midrule
program disturb Flash & minutes* & \makecell[l]{long response time} \\
\midrule
circuit-intrinsic & $\mu$s & \makecell[l]{safe and suitable voltage/frequency control} \\
\midrule
sensor & seconds & limited CRP numbers \\
\bottomrule
\label{tab:software-pufs-features}
\end{tabular}
\vspace{-9pt}
\begin{tablenotes}
    \footnotesize
    \item * The latency is not directly reported in~\cite{hasan2024nand, babu2026examination}
\end{tablenotes}
\end{threeparttable}
}
\end{table}

Other circuit-intrinsic PUFs are similar to processor-intrinsic PUFs, both using timing errors when working under stressed conditions, except that the entropy comes from pre-existing non-processor functional circuits. DSP-PUF~\cite{yuan2024dsp}, Yuan et al.~\cite{yuan2025overclocking} and SPUF~\cite{chen2021spuf} implemented PUFs on the DSP-28335, H.265 video encoding circuit, and encryption circuits, respectively. 
In addition, G-PUF \cite{forlin2020g} introduced a software PUF design using errors when a graphics processing unit operates under stressed conditions. 

\subsection{Flash and sensor PUFs}

Flash PUFs exploit device-specific program-disturb behavior in commodity NAND Flash. Hasan et al.~\cite{hasan2024nand} repeatedly programmed selected pages and used the resulting bit-flip patterns as responses. Babu et al.~\cite{babu2026examination} further evaluated this approach under thermal stress. These PUFs are runtime-accessible but require raw Flash access and wear-inducing operations. Sensor PUFs further extend the scope of software PUFs by leveraging ubiquitous sensors in IoT devices. Fukushima et al. \cite{fukushima2016sensor} implemented a PUF using only software APIs of the accelerometer and gyroscope. The challenge in their design is generated by shaking and twisting the hand wearing the device and the responses are the maximum and minimum values read from the sensors. Later, Fukushima et al. \cite{fukushima2019evaluation} expanded the idea by using only a gyroscope to implement a PUF. 

\subsection{Cross-Category Observations for Model Protection}

We evaluate the reported latencies, analyze the associated limitations and compare them according to the types of hardware components used to implement software PUFs, as well as the methods employed to induce stressed conditions or expose analog characteristics, as summarized in Table~\ref{tab:software-pufs-features}. The comparison shows that the suitability of a software PUF depends on the deployment setting. Start-up SRAM PUFs are suitable for key reconstruction during boot, but they cannot normally be queried at runtime. Most DRAM retention PUFs experience a fairly long response time compared to others, which would add directly to model-load time.
Among memory-based approaches, latency-based DRAM PUFs appear to be the most practical for runtime use because they generate responses within milliseconds. Their usage, however, depends on whether the platform allows control of DRAM timing parameters. Retention and rowhammer PUFs, whose response generation can take minutes, are better suited to one-time provisioning. Program-disturb Flash PUFs can be queried without rebooting, but require repeated writes, potentially causing substantial latency and wear. Sensor PUFs are easy to access through standard APIs on IoT and wearable devices, but their limited challenge diversity and dependence on physical motion make them unsuitable for model protection. Circuit-intrinsic PUFs can provide very low response latency, but none has been demonstrated on unmodified commercial processors. Overall, software PUFs for model protection should be selected based not only on uniqueness and reliability, but also on latency and runtime accessibility.

\section{Discussion}\label{discussion}

\textbf{Application of Software PUFs in solving model security concerns.} A scenario of how software PUFs solve this gap is to bind the model to a specific device without storing a key in non-volatile memory. The challenge-response pair is used to derive the key. For example, Jiang et al.~\cite{jiang2025puf} introduced a novel protection scheme by modulating the normalization block parameters with a credential built from PUF. This protection scheme enables the model to behave accurately only on authorized devices. Shao et al.~\cite{shao2024imce} apply the same principle at the architecture level. The cloud stores encrypted weights on the device and sends only challenges as masked keys. The on-chip PUF regenerates the actual key, so keys are never transmitted over unsecured channel and the weights are decrypted in place during computation. However, their design relies on a dedicated PUF circuit and cannot be deployed on already-fabricated devices. Many of the software PUFs listed in Table~\ref{tab:software-pufs} suit such AI deployment scenarios and can be deployed through firmware updates, provided they are selected with their limitations in mind, notably runtime accessibility and response latency.

\textbf{Challenges.} Despite the economic potential of enhancing security without requiring hardware replacement or redesign, we highlight four open design challenges of software PUFs in edge AI devices based on the analysis of the works in Table~\ref{tab:software-pufs}. 
\textbf{1) Exposing analog characteristics using only software APIs}. Most existing on-device edge AI devices offer diverse options for software PUF implementation, leveraging components such as DRAM, SRAM, microcontrollers, or GPUs. However, commodity digital devices are designed to hide their analog characteristics, modifying the firmware or implementing kernel modules to enable low-level access to the hardware might be necessary. In addition, the degree to which timing parameters should be skewed to induce stressed conditions while preserving normal system function remains a platform-specific calibration parameter. Edge SoCs often ship with closed firmware, so such low-level access may not be available.
\textbf{2) Preserving sufficient entropy during response extraction.} The challenge is not simply balancing uniqueness and reliability, since these metrics are primarily determined by the underlying entropy source and operating condition. Rather, the challenge is to design an extraction pipeline that preserves sufficient device-specific entropy while suppressing unstable response bits. Many designs mention fuzzy extractors, which use error-correction codes to translate the noisy response into a reproducible secret key. However, few studies~\cite{kietzmann2023puf, rahmati2015probable, fukushima2016sensor, fukushima2019evaluation} quantify the residual entropy after error correction or the information revealed by helper data. 
\textbf{3) Accessibility at runtime.} On-device AI deployment demands low latency to maintain fast responses. However, software PUFs often require disruptive conditions, such as system reset and overclocking. Operations such as system reset, which is required by SRAM PUFs, interfere with normal execution, which may cause severe delay in the model's response. By analyzing the runtime accessibility of the existing software PUFs and summarizing them in Table~\ref{tab:software-pufs}, we found that many DRAM PUFs and sensor PUFs are accessible, while most SRAM PUFs are not, as the response is only accessible at start-up. However, all DRAM PUFs, except for latency-based ones, require a long response time, making them less suitable for latency-sensitive real-world deployment. For model protection, response latency should be at least comparable to the model-inference time.
\textbf{4) Defining adversary models.} In on-device AI deployments, the adversary model must specify both the attacker's capabilities and goals, such as stealing the model, tampering with it, or causing denial of service. Because low-level software configures the hardware and obtains the response, an attacker who controls the operating system, kernel, firmware, or device driver may be able to invoke the PUF repeatedly, alter timing or voltage parameters, record and replay previous responses, or modify enrollment and helper data.

\textbf{Future research:} Future research should pursue three practical directions to advance software PUFs toward real-world deployment. First, parameter calibration should be automated on COTS platforms and systematically evaluated across different devices, operating conditions, and aging stages. Second, software PUFs should be evaluated against relevant attacks, including modeling attacks, reliability-based attacks using repeated measurements, and side-channel leakage from fuzzy-extractor implementations. Third, complete prototypes should integrate enrollment, response generation, and key reconstruction or authentication on unmodified commodity devices, while also reporting deployment metrics such as latency, service interruption, and effects on normal workloads.

\section{Conclusion}\label{sec:conclusion}

Edge AI devices remain in service for years because of the cost of updating. Protecting models on edge devices requires a device-bound root of trust, yet most deployed hardware cannot be redesigned to provide one. This survey examined software PUFs as a deployable alternative: PUFs that derive device-unique responses from existing components through software-level interfaces, with no design-stage hardware modification. We defined this class by accessibility, separating it from intrinsic PUFs, which are defined by entropy source, and from designs that need hardware additions. Using this definition, we reviewed 37 works across SRAM, DRAM, Flash, circuit-intrinsic, and sensor-based approaches, and compared their performance in terms of uniqueness and reliability, providing valuable insight into the trade-offs, challenges, and limitations. To the best of our knowledge, this is the first survey focused exclusively on software PUFs.



\bibliographystyle{IEEEtran}
\bibliography{bibliography}

\end{document}